\documentstyle[aps,preprint]{revtex}

\begin{document}

\title{ Computational Study of the Structure and Thermodynamic Properties of
Ammonium Chloride Clusters Using a Parallel J-Walking Approach }

\author{Alexander Matro and David L. Freeman}

\address{Department of Chemistry, University of Rhode Island \\
51 Lower College Road,
Kingston, Rhode Island 02881-0809}

\author{Robert Q. Topper}
\address{Department of Chemistry, University of Rhode Island \\
51 Lower College Road,
Kingston, Rhode Island 02881-0809 \\ and \\
Department of Chemistry, The Cooper Union for the Advancement
of Science and Art, Albert Nerken School of Engineering, 51 Astor Place,
New York, NY 10003$^*$}
\addtocounter{footnote}{1}
\footnotetext{Present Address}

\maketitle

\begin{abstract}
The thermodynamic and structural properties of (NH$_4$Cl)$_n$ clusters, n=3-10
are studied.
Using the method of simulated annealing, the geometries of
several isomers for each cluster size are examined.
Jump-walking Monte Carlo simulations are then used to compute the
constant-volume heat capacity for each cluster size over a wide
temperature range.
To carry out these simulations a new parallel algorithm is developed
using the Parallel Virtual Machine (PVM) software package.
Features of the cluster potential energy surfaces, such as
energy differences among isomers and rotational barriers of the ammonium ions,
are found to play important roles in determining the shape of the
heat capacity curves.
\end{abstract}
\pacs{PACS numbers: }

\pagebreak

\section*{I. Introduction}

As atomic or molecular aggregates ranging in size from 3
to as many as several thousand monomer units, clusters represent
an intermediate state of matter between the finite and
the bulk.  This intermediate character of clusters has led to
considerable interest in their
structural and thermodynamic properties.
>From the perspective of structure, it is frequently possible to determine
the potential energy minima,
transition states, and the general topography of the potential energy
surface.
The thermodynamic properties of clusters often mimic bulk
materials and have been shown to exhibit distinctive
``solid-like'' and ``liquid-like'' features.\cite{ffd1,ffd2,franz,beck,bjb}

The examination of the potential energy surfaces of clusters has
been a subject of numerous theoretical investigations.
A thorough study by Tsai and Jordan has uncovered many
of the isomers and transition states of (Ar)$_n$, $n$=7-13 clusters described
by a Lennard-Jones pair potential. \cite{tsai}
Studies of NaCl clusters and cluster ions by Phillips {\em et al.}
\cite{phillips}
have shown that clusters bound by ionic forces are capable of forming
a rich variety of structural isomers.
Complimentary to the study of potential minima is the study of structures
and energies of transition states connecting the minima.
In addition to the work of Tsai and Jordan \cite{tsai}, potential energy
surfaces of clusters have been explored for transition states
by Davis {\em et al.}~\cite{davis} and Wales. \cite{wales}
The height of the transition state barriers between isomers can play
a pivotal role in influencing the isomerization time scale in clusters.

Two approaches have been used to study the thermodynamic properties of
clusters.
The first approach \cite{bbdj,rb} uses Molecular Dynamics simulations to
identify temperature regions with liquid-like and solid-like behavior.
In the solid-like region clusters undergo small-amplitude
vibrations about their potential energy minima, while in the liquid-like
region large-amplitude motions are observed, taking the clusters
among different potential wells.
Lyndon-Bell and Wales \cite{lyndon} have used Molecular Dynamics
simulations of Lennard-Jones clusters to determine Landau free energy
barriers between solid-like and liquid-like states.

A second approach, used in this work, is to use Monte
Carlo simulations to calculate thermodynamic properties of the clusters such
as C$_V$, the constant-volume heat capacity.
Heat capacities for Lennard-Jones clusters \cite{ffd1,franz}
have been calculated in this manner,
and anomalies in C$_V$ such as peaks and shoulders have been found to
correspond to the onset of isomerization.
These isomerization transitions have been interpreted as
signatures of cluster analogs of phase changes for some cluster
sizes. \cite{franz}
In addition, Lopez and Freeman \cite{lopez} have calculated heat
capacities for Ni-Pd clusters, observing a ``melting'' transition and a
low-temperature anomaly corresponding to an order-disorder transition
analogous to those seen in bulk bimetallic alloys. \cite{bulkNiPd}

Metropolis Monte Carlo simulations\cite{mrrtt} in the temperature region
corresponding to the
onset of isomerization suffer from the inability to
sample all available configuration space ergodically.
The deficiency of Metropolis sampling, which cannot be effectively overcome by
increasing the step size, is its inability to move among the
statistically important regions of configuration space that are separated
by significant transition state barriers.
One way to overcome the problems of Metropolis sampling to use the the
jump-walking (or J-walking) move strategy.\cite{ffd1,ffd2}
J-walking combines the small step size of Metropolis Monte Carlo with
occasional jumps to configurations belonging to a higher temperature
ergodic distribution.
The higher temperature distribution contains information about the potential
energy surface of the system, and the jumps to configurations belonging to
separated but statistically important regions of configuration space
are thus attempted with sufficient frequency.

Additional complexities in cluster behavior can occur in molecular aggregates.
For example, in the case of ammonium chloride studied here,
the effects from the rotations of the ammonium ions significantly
contribute to the thermodynamic properties.
The rotational potential energy barriers are, in many instances, smaller than
barriers between distinct minima, causing the rotational effects to
be reflected in the heat capacity at lower temperatures than the
isomerization effects.

In the current work we report two principal outcomes.
First, we develop a parallel
J-walking algorithm that enables us to carry out a Monte Carlo
simulation efficiently in a multi-processor computing environment.
Second, the parallel J-walking algorithm is applied to the study of the
thermodynamic properties of (NH$_4$Cl)$_n$, $n$ = 3-10 clusters.
The contents of the remainder of this paper are as follows.
In Section II we describe the computational methods,
including the model potential and a discussion of the simulated
annealing methods used to locate the ammonium chloride cluster isomers.
This is followed by a discussion of the J-walking Monte Carlo simulation
technique used in computing the constant-volume heat capacity, and the
implementation of Parallel Virtual Machine (PVM) software \cite{pvm} into our
Monte Carlo algorithm.  Section II concludes with a discussion of the
methods used in locating transition states.
In Section III we present our results.  Isomers of the ammonium
chloride clusters found by simulated annealing are shown, along with
a few structures of transition states.  Then, a constant-volume heat
capacity curve for each cluster size is presented, along with a
discussion of the features of each curve.  Finally, Section IV contains
our concluding remarks and directions for future work.

\section*{II. Methods}

\section*{A. Model Potential}
In the Monte Carlo simulations carried out in this work, a pair
potential dominated by Coulombic interactions is used.
The form of the potential is
\begin{equation}
V({\bf r}) = \sum_{i}\sum_{j>i} ( A_{ij}\exp{(-\alpha_{ij}r_{ij})}
+\frac{q_iq_j}{r_{ij}}
+\frac{D_{ij}}{r_{ij}^{12}} - \frac{C_{ij}}{r_{ij}^{6}} )
\label{eq:potential}
\end{equation}
where ${\bf r}$ represents the entire set of coordinates for the system,
$r_{ij}$ is the distance between particles $i$ and
$j$, $q_{Cl}=-1.0$, $q_N=-0.4$, $q_H=0.35$, and the remaining
parameters are given in Table 1.
The potential is derived from potentials
of Klein {\em et al.} \cite{klein}
and Pettitt and Rossky \cite{pr1,pr2}.
The parameters not available from other sources
are obtained with the standard combination rules.
Because the present simulations are classical and the internal
vibrations of the ammonium ions are expected to be quantum mechanical
with high frequencies,
the ammonium ions in the potential of Eq. (\ref{eq:potential}) are
assumed to be rigid tetrahedra.
With the potential in Eq. (\ref{eq:potential}), in the CsCl phase
at 0K the
lattice constant for bulk ammonium chloride
is found to be 3.79 \AA \ with a cohesive energy of
-720 kJ mol$^{-1}$.  These numbers can be compared with the
experimental lattice constant (3.868 \AA)\cite{lub} and the experimental
cohesive
energy at 298K (-697 kJ mol$^{-1}$).\cite{wil}

Because clusters have finite vapor pressures in constant
temperature simulations, an external constraining potential\cite{abe} has been
included about the center of mass of each cluster.  For this constraining
potential we have chosen the same form used elsewhere\cite{fclu}
\begin{equation}
V_c({\bf r}) = \kappa \sum_{i=1}^n \left ( \frac{|{\bf r}_i-{\bf R}_{cm}|}{R_c}
\right )^{20}
\end{equation}
where {\bf r}$_i$ is the coordinate of particle $i$, {\bf R}$_{cm}$ is
the coordinate of the center of mass of the cluster, $\kappa$ has units
of energy and $R_c$ is a
parameter that defines the radius of the constraining potential.  In the
current calculations $R_c$ is taken to be 15 Bohr for the trimer, 20
Bohr for the tetramer, 25 Bohr for $n=5-7$, and 30 Bohr for $n=8-10$.
In this work
$\kappa$ is taken to be unity.

In the simulated annealing calculations that are used in finding
structures of the isomers, it is computationally more convenient to
include all degrees of freedom of the system.
The internal vibrations of the ammonium ions are described with a
harmonic force-field potential of the form
\begin{equation}
V_I = \frac{1}{2}k_{NH}\sum_{i=2}^{5}(r_{1j}-r_0)^2 +
\frac{1}{2}k_{\theta}\sum_{i=2}^{4}\sum_{j=i+1}^{5}(\theta_{ij}-\theta_0)^2
\label{eq:ammpot}
\end{equation}
where atom 1 is the nitrogen, and atoms 2-5 are the hydrogens,
$\theta_{ij}$ is the angle formed by hydrogen atoms $i$ and $j$ and the
nitrogen atom, $\theta_0$ is the H$-$N$-$H angle in a perfect tetrahedron,
and $r_0$=1.9467 Bohr.
The force constants, $k_{NH}$=0.351 Hartree/(Bohr)$^2$ and
$k_{\theta}$=0.137 Hartree, are given by
Herzberg. \cite{herzberg}
The harmonic force-field potential for NH$_4^+$ yields vibrational
frequencies of 1493, 1692, 3045, and 3193 cm$^{-1}$.
These values compare favorably with the experimental frequencies of
1398, 1699, 3047, and 3129 cm$^{-1}$. \cite{harvey}

\section*{B. Simulated Annealing}

The search for the isomers of the NH$_4$Cl clusters
has been carried out using the method of simulated annealing.
This method has been used previously to locate the isomers of
other systems.
The basic idea of simulated annealing is to take a system at
high temperature and gradually cool the system until the
global minimum or a local minimum on the potential energy surface is
attained.
If the cooling is performed adiabatically, the system will find the
global minimum.
Using a finite cooling rate allows the system to become trapped in local
minima, yielding information about the various isomers.

In this work we use a Brownian dynamics approach. \cite{lopez,chandra,annea}
Starting with chloride and ammonium ions randomly distributed in a fixed
volume, the system is propagated in time according to the Langevin equation.
As the temperature is decreased, the kinetic energy is drained from the system.
When the temperature finally reaches 0 K, the friction term in the
Langevin equation drains the remaining kinetic energy from the system,
leaving it in some local minimum.

For the smaller clusters the number of isomers is manageable,
and the lowest energy isomer for each cluster size can be identified
from the Brownian dynamics simulations.
For the larger clusters, however, the number of isomers becomes
overwhelming, making it increasingly difficult to identify the lowest
energy isomer.
We use the J-walking Monte Carlo simulations described below to
confirm that the lowest energy isomer is correctly identified.
Because the J-walking Monte Carlo simulations are expected to be fully ergodic,
the cluster will be in its lowest energy isomer as the
simulation temperature approaches 0 K.

Unlike the Monte Carlo simulations, we do not assume rigid ammonium ions
in the Brownian dynamics simulations.
In most cases
the differences between the structures determined by Brownian
dynamics and the corresponding rigid-ammonium structures are
small.  The main differences consist of slight distortions of the
ammonium ions away from tetrahedral geometries and slight distortions
of the entire cluster.
To find the rigid-ammonium structures from their fully relaxed counterparts,
we take the fully relaxed structure and carry out an additional
Brownian dynamics simulation at 0 K.
During this simulation, force constants
$k_{NH}$ and $k_{\theta}$ in Eq. ({\ref{eq:ammpot}) are gradually
increased until the N$-$H bond distances and H$-$N$-$H angles are
almost at their equilibrium values.
Next, the ammonium ions are replaced with perfect tetrahedra, and the
energy is further minimized with Monte Carlo moves.
All isomers that we have found for (NH$_4$Cl)$_3$ and (NH$_4$Cl)$_4$,
along with the lowest energy isomers for the remaining cluster sizes
are presented in Section III.

\section*{C. J-walking}

It has been seen in several systems that if isomers
are separated by high potential energy barriers, Monte Carlo
simulations using the ordinary Metropolis move strategy \cite{mrrtt} can result
in
non-ergodic sampling of configuration space, leading to large errors in
calculated averages.
This problem can be most acute in the intermediate temperature range
at the onset of isomerization.
Calculated quantities that suffer most severely from the non-ergodicity of
Metropolis Monte Carlo are the quantities involving energy fluctuations, such
as the constant-volume heat capacity, C$_V$, given by
\begin{equation}
\frac{C_V}{k_B} = \frac{3}{2}n +
\frac{\langle V^2 \rangle - \langle V \rangle^2}{(k_BT)^2}
\label{eq:cv}
\end{equation}
where $\langle V^2 \rangle$ and $\langle V \rangle^2$ are the average of
the square and the square of the average of the potential energy,
respectively, $T$ is the temperature, $n$ is the number of particles,
and $k_B$ is the Boltzmann constant.

The non-ergodicity of the random walk can be alleviated using a move
strategy called J-walking.
J-walking combines ordinary Monte Carlo moves
with jump attempts to configurations in an ergodic distribution at a
higher temperature.
We discuss briefly J-walking here and refer the reader to original papers
for detailed descriptions.\cite{ffd1,ffd2}

For clusters, ordinary Metropolis sampling is normally adequate only for
high and low temperatures.
At high temperatures, the system is in a fluid-like state, and
the combination of high temperature and large Monte Carlo step size is
sufficient to overcome barriers among the potential energy wells.
On the other hand, only the global minimum is thermally accessible
at low temperatures and Metropolis sampling about the global
minimum is sufficient.
The intermediate temperature range presents a problem for Metropolis
sampling because the barriers separating the cluster isomers can result in
transitions between potential minima with insufficient frequency.

The main idea behind J-walking is to use information about the
potential surface obtained in a high temperature simulation.
A walker at a lower temperature executes an
ordinary Monte Carlo walk,
occasionally attempting jumps to configurations at the higher temperature
where the walk is ergodic.
Each distinct isomer is sampled adequately with the ordinary Metropolis
move strategy,
and the occasional jumps to the high temperature distribution ensure
that all isomers are included with the proper frequency.
Details can be found in the original literature.\cite{ffd1,ffd2}

In practice, J-walking can be implemented in two ways, each with
its own set of drawbacks.
The first approach uses tandem walkers, one at a high temperature
where Metropolis sampling is ergodic,
and one or multiple walkers at lower temperatures.
The configurations generated by the high temperature walker are used
by the lower temperature walkers for attempting J-walking moves.
This tandem approach has been used infrequently because correlations
inherent in the Metropolis walks can introduce systematic errors in the
J-walking results.

The second approach writes the configurations from the simulation
at the J-walking temperature to an external file and accesses
configurations randomly from the external file while carrying
out a simulation at the lower temperature.
Correlation errors are avoided with external configuration
files because of two features in this approach.  First,
as discussed in Reference \cite{franz}, by writing configurations to an
external file infrequently (about once every 40-100 moves), the correlations
are partially broken.  Second, because
correlations still persist between configurations separated
by 40-100 moves, it is also necessary to access the external files
randomly.
The difficulty with this approach is the storage requirements
for the external distributions.
The large storage requirements
have limited the application of the method only to small systems.

\section*{D. Parallel Strategy}

As we discussed above, J-walking can be implemented using either
tandem walkers or previously generated external distributions.
The best features of these two approaches can be combined into a single
J-walking algorithm with the use of multiple processors and the Parallel
Virtual Machine (PVM) software package. \cite{pvm}
PVM enables processes running on the same or different
machines to exchange information while they are executing.
We incorporate the PVM subroutines into the Monte Carlo
computer code, where these subroutines are used to send and receive
configuration geometries and potential energies of the clusters.

Using a multiple-processor computing environment, we can execute a
separate Monte Carlo walk on each processor, with PVM enabling
these Monte Carlo walks to communicate with each other.
Instead of generating external distributions and storing them
before the actual simulation,
we generate the required distributions ``on the fly'' and pass them to
the lower-temperature walkers.
The walker accepts or rejects the configuration it receives based on
Eq. (5) of Reference \cite{ffd1}.

Our computational scheme consists of two kinds of Monte Carlo processes,
the configuration generating processes and the computing processes.
The generating processes are designed to be a source of ergodic
distributions at particular temperatures.
Each set of computing processes carries out Monte Carlo walks at
temperatures below that of the generating process, periodically jumping
to a configuration provided by the generating process to maintain ergodicity.

A diagram of a model PVM J-walking simulation is shown in Fig. 1.
Each box represents a process in a parallel machine executing a Metropolis
simulation at a particular temperature.
The set of boxes on the left-hand side of Fig. 1 represents the
generating processes, executing walks at temperatures
T$_1$, T$_2$ and T$_3$.
T$_1$ is assumed to be sufficiently high that Metropolis Monte Carlo is
ergodic without modification.
The set of boxes on the right-hand side of Fig. 1 represent the
computing processes.
The necessity of having the generating processes running at different
generating temperatures is precisely analogous to generating external
distributions at these temperatures in a J-walking simulation
using the serial algorithm.\cite{ffd1}
As in the serial J-walking method, a new distribution is needed
when the jump acceptance rate falls below 10-15\%.
The generating processes at T$_2$ maintain their ergodicity just like
the computing processes by making jump attempts to configurations
contained in the T$_1$ processes.
In turn, the T$_2$ processes serve as sources of ergodic distributions
for computing processes executing at temperatures between T$_2$ and T$_3$.
Additionally the generating processes at T$_3$ are made ergodic by
periodic jumps to the configurations contained in the T$_2$ processes.
If still lower temperature generating processes are needed, this
procedure can be continued.
It is important to note that four processes are executed at each generating
temperature.
The four downward arrows from T$_1$ to T$_2$ and from T$_2$ to T$_3$
denote that the generating processes at each temperature
are independent of each other.
The boxes on the right-hand side of Fig. 1 represent the computing
processes.
The thick arrows pointing from the
generating processes to the computing processes indicates that each
of the
four generating processes is feeding configurations to each
computing process.

The reason for having four generating processes at each temperature
instead of just one is the following.
The bottleneck of this algorithm is the frequency at which a
generating process can pass a configuration to a computing
process or to a lower-temperature generating process.
This frequency must be kept sufficiently low to allow the Monte Carlo
simulation at the generating temperature ample opportunity to alter the
configuration of the system.
Previous J-walking studies of Lennard-Jones clusters have shown that a
configuration can be saved once every 40-100 Monte Carlo passes,
depending on the size of the cluster.\cite{franz}
It has also been determined that a computing process should
attempt jumps to the
ergodic distribution as often as once every 10-15 passes.
We can maintain the 10-15 pass criterion in the computing
process by having each computing process access four independent generating
processes.  The result is that each generating process is accessed once
every 60 moves, but since there are four, the computing process
can attempt a jump to one of the four generating processes once every 15 moves.
In tests performed on (NH$_4$Cl)$_9$, storing a configuration every 100
passes instead of every 60 did not affect the calculated heat capacity.
The computing time is reduced by a factor of four in the case
where the number of processors is not limited.

As in the serial algorithm there are correlations in the Metropolis walks of
each generating process that must be broken.
To break the correlations, each configuration-generating process
keeps an array of configurations in memory.
When a configuration is transmitted to a lower-temperature process,
it is a configuration randomly chosen from this array, not the current
configuration of the walker.
The current configuration of the walker then replaces the configuration
just passed from the array to another processor.
The minimum size for this array has been found to be 2500 configurations
from tests on the (NH$_4$Cl)$_9$ cluster.  In all calculations presented
below, a 5000-configuration array is used.
Compared to array sizes of at least 50000 configurations used in
the serial algorithm, the arrays in the parallel method are small and do not
inhibit applications of the method to large systems.

An issue that must be addressed is the initiation of the
5000-configuration array at each generating temperature T$_1$, T$_2$, etc.
We have examined two approaches.
One approach is to create the arrays during the computation.
The startup time required to populate the
configuration arrays sequentially can be prohibitively long.
For example, the computing processes between T$_1$ and T$_2$ and the generating
processes at T$_2$ must wait until the 5000-configuration arrays
at T$_1$ are created.
The other approach, which we have found to be more useful, is to
create the small 5000-configuration
distributions using serial J-walking prior
to the PVM calculation.  The initiating walks are short and require
storage of arrays of only 5000 configurations.
This approach allows the most efficient use of the multi-processor
computing environment because there are no idle processors at any time
during the PVM calculation.

For the computations in the present work
the typical number of Monte Carlo passes for each temperature point
is 1.5 million for cluster sizes 4-10.
During each run, 10$^5$ J-walking attempts
are made.  The calculations for the (NH$_4$Cl)$_3$ cluster consist of
3 million moves for each run and 5$\times$10$^4$ J-walking attempts.
A typical PVM J-walking simulation for the larger ammonium chloride
clusters requires a total of about 90 processes to span the
entire temperature range of interest.  For these calculations, 45
individual processors are used, with two processes executing on each processor.

\section*{E. Locating Transition States}

Although the energies and geometries of local minima on the
multi-dimensional potential energy surface of a cluster are important
determining factors for the onset of isomerization, a crucial role is
also played by the height of the potential energy barriers separating
the minima.
Determining the height of these barriers entails the search for the
transition states on the potential energy surface.

We use two methods to search for transition states.  The first method
is an approach previously developed to find the transition
states of Lennard-Jones clusters. \cite{mfd}
Briefly, transition states are
located by constructing double-ended classical trajectories that connect a
pair of local minima.
The trajectories are located by expressing the
path between two isomers of interest with a Fourier expansion about
a constant-velocity path.
The Fourier coefficients are then determined, yielding a trajectory with
a particular total energy.  The details of finding the Fourier coefficients
are given elsewhere.\cite{mfd,dbf}
The trajectory with the lowest total energy is usually found to
pass through or near the transition state, identified by having
exactly one negative eigenvalue in its Hessian matrix.

The other method used in locating the transition states is the
eigenmode-following technique initially proposed by Hildenbrand
\cite{hildenbrand}
and developed further by other workers. \cite{cerjan}
More recently, Tsai and
Jordan \cite{tsai} and Wales {\em et al.} \cite{wales} have employed
this method to locate
transition states of atomic and molecular clusters.
Briefly, a search is conducted by starting at a local minimum and
following one of the normal modes by maximizing the potential energy
along that particular mode while minimizing the potential energy along
the remaining normal modes, until a transition state is reached.

For the case of (NH$_4$Cl)$_4$ and (NH$_4$Cl)$_9$ we are concerned with
rotational barriers of the ammonium ions.  The double-ended trajectory
method for finding transition states is ideally suited to these motions
because of their simplicity.
For the more complex transition states of the (NH$_4$Cl)$_3$ cluster
connecting distinct
isomers we use the eigenmode-following method.
The double-ended trajectory method is unable to
locate the transition states between the isomers of the trimer with a low
number of Fourier expansion coefficients.
To apply the double-ended trajectory approach to a case with a
complex rearrangement, a large number of Fourier coefficients is
required to describe the path of the system between the two
minima, making the method computationally expensive.
Rather than pursuing the transition state search with the double-ended method,
we
have opted to use the eigenmode-following method in the transition
state searches in (NH$_4$Cl)$_3$.

\section*{III. Results}
\section*{A. Structures}

The ammonium chloride monomer is known \cite{howard} to be a van der Waals
complex
rather than an ionic species and cannot be treated using the
model potential of Section II.A.
We do not investigate (NH$_4$Cl)$_2$ because it has only a single isomer
and is not expected to exhibit interesting thermodynamic behavior.
We assume that all clusters in this work are ionic.
The cluster size at which there is a transition from van der Waals to
ionic bonding represents an interesting and complex electronic structure
problem that is beyond the scope of this work.

We begin our discussion of the  ammonium chloride cluster structures
with (NH$_4$Cl)$_3$, for which we have found three isomers with our
potential model.
These three isomers, located with the simulated annealing
procedure described in the previous Section, are shown in Figs. 2(a),
2(c) and 2(e).  Structures (b) and (d) in Fig. 2 are the transition states
connecting (a) with (c) and (a) with (e), respectively.
The transition states shown in Figs. 2(b) and 2(d) are determined using
the full potential, including the internal vibrations of the ammonium
ions.
Because our transition state search methods determine the transition states
using the full potential, all transition state barriers reported in this
work are calculated relative to the potential energy minima of the full
potential.
The corresponding transition state barriers for the structures with
rigid ammonium ions are expected to be roughly the same.
Interestingly, isomers (a) and (c) are so close to each other in energy
that isomer (c) becomes the lowest energy isomer if the full potential is
used and the ammonium ions are allowed to relax fully.

Figures 3(a) - 3(f) show the isomers we located for the (NH$_4$Cl)$_4$ cluster
in the order of increasing potential energy.
There is a qualitative difference in structure between the lowest isomer (a)
and the remaining isomers (b)-(f).
The lowest energy isomer has a compact rock-salt structure while the remaining
isomers have open-type structures.
Unlike (NH$_4$Cl)$_3$, in which all three isomers are
close to each other in energy, there is a 13,485 K gap separating
the two lowest lying isomers of the tetramer, (a) and (b).

Figure 4 shows the lowest energy isomers for clusters sizes 5-10.
It is interesting to note the trend in the growth pattern for the
clusters with an even number of ion pairs.
Beginning with (NH$_4$Cl)$_4$ (see Fig.3(a)), each successive
even-number
cluster is formed via the addition of an identical (NH$_4$Cl)$_2$ unit.
The odd-number clusters do not show a pattern of formation from
either the smaller odd-number cluster or the preceding even-number cluster.

The relative stability of the even-number clusters versus the odd-number
clusters can be compared by examining the gain in binding energy of the
lowest energy isomer, $\Delta V = V_{n-1}-V_n$ as a function of the
cluster size $n$, where $V_n$ is the potential energy of the lowest
energy isomer of size, $n$.
The trend observed in Fig. 5 clearly shows that the gain in potential energy is
greater when going from an odd to an even cluster, except for
(NH$_4$Cl)$_9$.
The deviation of (NH$_4$Cl)$_9$ from the observed trend
can be understood by examining its geometry, shown in Fig. 4(e).
The (NH$_4$Cl)$_9$ cluster forms a rock-salt structure that
has more in common with the even-number clusters than with the other
odd-number clusters.

\section*{B. Heat Capacities}

The constant-volume heat capacity curve for (NH$_4$Cl)$_3$ is
shown in Fig. 6.
The error bars in this and all subsequent heat capacity curves represent
two standard deviations.
Also, only the potential energy contribution to the heat capacity is
shown in this and all subsequent heat capacity curves, i.e. we set
\begin{equation}
C_v = \frac{\langle V^2 \rangle - \langle V \rangle^2}{(k_BT)^2}.
\end{equation}
There are no additional structural features from the constant kinetic
energy contribution in Eq. (\ref{eq:cv}).
The curve starts at approximately 10.5, the equipartition value for the
trimer, undergoes a slow increase in the 0-80 K range, then rises
rapidly to a peak at 140 K, and finally the heat capacity declines gradually
over the remaining temperature range.
The presence of an early peak in the heat capacity is a direct
consequence of the small differences in energy among the
isomers.

We have found that during the simulation at 50 K all
configurations are in the potential energy wells belonging to isomers
(a) and (e) in Fig. 2, and none are found in the well belonging to
isomer (c).
At 250 K, less than 0.03\% of configurations are found in the well of
isomer (c).
The isomers occupied during a finite-temperature simulation
are identified using the nitrogen-chloride distances and the
angle between the planes defined by the nitrogens and the chlorides.
Since the absence of structures belonging to the potential energy well
of isomer (c) is clearly not a consequence of energetics, we have determined
transition state barriers between isomers (a) and (c) and between isomers
(a) and (e).
These transition state geometries are shown in Figs. 2(b) and 2(d),
respectively.
As discussed in Section III ~A and listed in Fig. 2,
the transition state barriers in both
cases are about 2000K above isomer (a).
The difference in the potential energy barrier heights
does not appear to be the reason for the extremely rare occurrence
of isomer (c) in our simulations.
One possible explanation for the small contribution of
structures belonging to the potential well of isomer (c) may be the
small fraction of
configuration space available to the isomer.
A normal mode analysis of isomers (a) and (c) has revealed that the lowest
vibrational frequency of isomer (a) is approximately one-half of the
the corresponding frequency of isomer (c).  Furthermore, the lowest
vibrational frequency of isomer (e) is one-third of the corresponding
frequency of isomer (c).
The frequencies of the first six vibrational modes of isomer (a) are 36,
41, 57, 67, 86, and 94 cm$^{-1}$. For isomer (c) these are 74, 76, 107,
150, 159 and 163 cm$^{-1}$, and for isomer (e), the frequencies are
26, 26, 36, 42, 42 and 54 cm$^{-1}$.
Although this difference in normal mode frequencies is not a direct
measure of the relative fraction of configuration space occupied by
the respective isomers, it is nonetheless a reflection of the narrowness of the
potential energy well of isomer (c) relative to isomers (a) and (e).

The heat capacity curve for the (NH$_4$Cl)$_4$ cluster, shown in Fig. 7, is
significantly different from that of (NH$_4$Cl)$_3$ in Fig. 6.
The peak in the (NH$_4$Cl)$_4$ heat capacity occurs at 1100 K.
The heat capacity increases slowly until 800 K, at which point
it rises rapidly to the aforementioned peak at 1100 K.
By examining the structures of (NH$_4$Cl)$_4$ during the simulations, we
have found that this large peak is a consequence of isomerization
transitions to the open structures shown in Fig. 3.
For (NH$_4$Cl)$_4$ and the remaining clusters we shall define the
``melting transition region'' to be the range of temperatures where
isomerizations to such open structures take place.
Another feature of the heat capacity curve shown in Fig. 7 is its
almost flat slope in the 400-700 K region.
This particular feature is present in most
of the cases studied in this paper.
As we shall discuss in the context of (NH$_4$Cl)$_9$,
the temperature region where C$_v$ is flat is the region where
the NH$_4^+$ ions are nearly freely rotating.
Using the method of double-ended classical trajectories, the barrier
for rotation of a NH$_4^+$ ion about the N-H bond pointing outward from the
cluster has been determined to be 4300 K.

The heat capacity curve for (NH$_4$Cl)$_5$ is shown in Fig. 8(a).
It has qualitative similarities to the (NH$_4$Cl)$_4$ heat capacity
shown in Fig. 7, with a melting peak in the 1000-1100 K region.
The melting peak is, however, significantly smaller relative to the
T = 0 K heat capacity than in (NH$_4$Cl)$_4$.
This is a trend that we observe for all the odd-number clusters, with
the exception of (NH$_4$Cl)$_9$; namely the ratio of the maximum
heat capacity relative to the T = 0 K value is significantly larger for the
even-number clusters than for odd-number clusters.

Unlike the tetramer, (NH$_4$Cl)$_5$ has several low-energy isomers.
Isomerization in (NH$_4$Cl)$_5$ is seen at significantly lower
temperatures than in the tetramer.
The isomerization at the lower temperatures in (NH$_4$Cl)$_5$ occurs
among isomers with compact structures, two of which are shown in Fig. 9
(a) and (b).  Figures 9(c) and 9(d) show examples of higher-energy open
structures of (NH$_4$Cl)$_5$.
The open-type structures are at least 12000 K above the lowest-energy isomer
and become accessible in the vicinity of the 1000-1100 K melting peak.
In contrast, only the lowest energy isomer of (NH$_4$Cl)$_4$ has a compact
structure and the remaining isomers are all high energy open structures
(see Fig. 3).

Figure 8(b) shows the heat capacity curve for (NH$_4$Cl)$_6$, and the
lowest energy isomer is shown in Fig. 3(b).
The heat capacity curve is similar to that of (NH$_4$Cl)$_4$.
The region between 500 K and 800 K is remarkably flat, and the melting
peak at 1200 K is sharp compared with (NH$_4$Cl)$_5$.
(NH$_4$Cl)$_6$ has at least one other low-energy isomer (not shown).
These low-energy isomers of (NH$_4$Cl)$_6$ are 2051 K apart and
isomerization between them is seen at temperatures as low as 400 K.
The remaining isomers found in this work are open-type structures and are
in the range of 12000-14000 K above the lowest energy isomer,
becoming accessible at temperatures near the melting peak.

The lowest energy isomer for the (NH$_4$Cl)$_7$ cluster is shown in Fig. 3(c).
The heat capacity curve, shown in Fig. 8(c), is the only curve in this
study that clearly exhibits two distinct peaks.
The first peak occurs around 600 K and is in the same region where other
cluster sizes (most notably 4,6,8, and 9) show a flat region.
In this case, however, the peak at 600 K is a result of isomerization
between the global minimum and a local minimum approximately 2000 K
above the global minimum.
The second peak, centered at about 1200 K, is a melting peak indicating the
onset of isomerization to open-type structures.

To investigate further the origin of the 600 K peak in the heat
capacity of (NH$_4$Cl)$_7$, we have performed quench studies of
the configurations
obtained from the simulations at temperatures of 425, 600, and 800 K.  At each
temperature, configurations were taken each 750 Monte Carlo passes, and
400 configurations at each temperature were quenched to the nearest
local minimum using Brownian dynamics.
At 425K, well before the peak in the heat capacity, almost all 400
configurations from an ergodic distribution
quench to the lowest energy isomer.
The dominance of the lowest energy isomer is reflected in the
large peak in the histogram in Fig. 10(a)
at 0 K and the small peaks representing higher energy isomers (the energies
of the isomers in Fig. 10 are given relative to the lowest
energy isomer defined to be 0 K).
At 600 K (Fig. 10(b)) slightly more than half of the configurations
quench to the lowest energy isomer, with the majority of remaining
configurations distributed between two isomers 830~K and 2194 K above
the lowest energy isomer.
The time-scale for isomerization at 600 K
is still slow, corroborated by examining configurations of
the (NH$_4$Cl)$_7$ cluster saved during the simulation.
The infrequent isomerization events result in large fluctuations in the
potential energy, leading to a peak in the heat capacity at 600 K.
Finally, at 800 K (see Fig. 10(c)) two isomers are seen to dominate the
histogram with roughly equal populations.
Rapid isomerization near 800 K leads to the observed
decrease in the heat capacity.
The second peak at 1200 K is the melting peak.

To determine the features of (NH$_4$Cl)$_7$ that are unique in this
series, we compare the energetics of the clusters.
(NH$_4$Cl)$_4$ has only one
compact structure, the global minimum (see Fig. 3), and the rest of the
isomers are at least 13000 K higher in energy.
(NH$_4$Cl)$_5$ has a few low-energy isomers (see Fig. 9), but there
is a large gap between the group of low-energy isomers and the group of
open-type higher energy isomers.
(NH$_4$Cl)$_6$ has just one other low energy isomer in addition to the
global minimum, and the energy difference between the two low energy
isomers in (NH$_4$Cl)$_6$ is roughly the same as the difference between
the two isomers of (NH$_4$Cl)$_7$ that dominate the the histogram in
Fig. 10(c).
However, in contrast to the case of (NH$_4$Cl)$_7$, (NH$_4$Cl)$_6$
does not have any other low-energy isomers.
The melting peak in (NH$_4$Cl)$_6$, as in the other even-number
clusters, is significantly larger compared to the T = 0 K heat capacity
than in the odd-number clusters.
Any decline in the heat capacity accompanying the
onset of rapid isomerization between the two low-lying isomers of
(NH$_4$Cl)$_6$ is masked in the 800 K region by the
melting peak.
Conversely, the melting peak in the heat capacity of (NH$_4$Cl)$_7$
is small and does not mask the drop in the heat capacity.

The structure (Fig. 3(d)) and the heat capacity of (NH$_4$Cl)$_8$ (Fig. 7(d))
are similar to those of (NH$_4$Cl)$_6$ (see Figs. 3(b) and 7(b),
respectively).
The slope of the curve in the 500-800 K ``shoulder''
region is not as flat as in the case of (NH$_4$Cl)$_6$.
Similar to (NH$_4$Cl)$_6$, there are a few low-energy isomers with
compact structures accessible at lower temperatures, and the majority of
isomers are higher-energy open-type structures that become accessible in
the vicinity of the melting peak.

The (NH$_4$Cl)$_9$ cluster, shown in Fig. 3(e), has more in common with
the even-number clusters than it does with the odd-number clusters.  The
structure of (NH$_4$Cl)$_9$ is slab-like, reminiscent of rock-salt
structure of NaCl.
The heat capacity curve, shown in Fig. 7(e), shows a flat region in the
400 - 800 K region and a large melting peak at about 1100 K.
An interesting feature of the (NH$_4$Cl)$_9$ cluster is the
low rotational barrier of the ammonium ion in the center of the cluster.
The center ammonium ion has 12 identical potential energy minima as
it rotates by 360 degrees, and the barrier to rotation between each pair
of minima is only 48 K.
Although we can see the center ammonium ion hopping between different
minima at temperatures as low as 5 K, the contribution of this motion to
the heat capacity is small and masked by the steady rise in the
heat capacity from the anharmonic vibrational motions in the
rest of the cluster.

We have claimed that the main
contributing factor to the appearance of a temperature region
where the slope of C$_v$ is small is the
onset of free rotation of the ammonium ions.
As a qualitative test of
this claim we carry out two additional Metropolis Monte Carlo simulations
in this temperature range for (NH$_4$Cl)$_9$.
In the first simulation, only the rotational Monte Carlo moves are
allowed, with the translational moves of the ammonium and chloride ions
excluded from the simulation.
In the second simulation, the translational moves of the ammonium and
chloride ions are included, but the rotational moves of the ammonium
ions are not allowed.
It must be stressed that this calculation has been done to
demonstrate qualitatively the effects of the onset of free rotation on
the heat capacity.
The two curves are shown in Fig. 11.
The upper curve
is the heat capacity resulting from the translational moves, and the
lower curve includes only rotations.
The upper (translational) curve increases through the entire
temperature range in Fig. 11.
The lower (rotational) curve rises to a maximum at about 600 K
and then undergoes a decrease.
The increase in the rotational contribution to the heat capacity
coincides with the ammonium ions undertaking infrequent hops between the
equivalent orientations.
Above 600 K, however, the motions of the ammonium ions are more
representative of free rotation rather than hops, which is consistent with
the decreasing rotational contribution to the heat capacity.
We have confirmed this by examining the
configurations of the (NH$_4$Cl)$_9$ cluster at several temperatures
throughout the flat region of the heat capacity.
The rotational barrier of a corner ammonium ion is 2908 K, and the
rotational barrier of a side ammonium ion is 1948 K.  Both of these
values are consistent with the onset of free rotation at 600 K.
The sum of the rotational and translational contributions is nearly flat as in
the full C$_v$ curve shown in Fig. 8(e).

The heat capacity for the remaining cluster in the study, (NH$_4$Cl)$_{10}$,
is shown in Fig. 8(f) and the lowest-energy isomer is shown in Fig. 3(f).
The heat capacity rises steadily, reaches its maximum at 700 K and then
slowly decreases with increasing temperature.
There is a change in slope in the 1000 - 1200 K region, where clusters of
size 4-9 exhibit a well-defined peak.

\section*{4. Concluding Remarks}
In this paper we have presented a study of structures and
temperature-dependent thermodynamic properties of (NH$_4$Cl)$_n$
clusters, $n$=3-10.  The structures have been determined using the approach of
simulated annealing, and the constant-volume heat capacities have been
computed using a newly developed parallel PVM J-walking algorithm.

The lowest energy isomers of the even-number clusters
form similar slab-like structures (see Figs. 3(a), 4(b), 4(d), and
4(f)).
The lowest-energy isomers of the odd-number clusters
(see Figs. 2(a), 4(a), 4(c), and 4(e)) generally do not show a clear
similarity with either the preceding even-number cluster or the previous
odd-number cluster.

We can compare our results to the structures of (NaCl)$_n$
clusters determined by Phillips {\em et al.} \cite{phillips}
For (NaCl)$_3$, only one isomer is found, compared to three for
(NH$_4$Cl)$_3$.  The structure of (NaCl)$_3$ is a planar ring, similar
to the highest energy isomer of (NH$_4$Cl)$_3$, shown in Fig. 2(e).
A close examination of the (NH$_4$Cl)$_3$ isomer shown in Fig. 2(c)
together with the absence of such a structure in (NaCl)$_3$ suggests
that the (NH$_4$Cl)$_3$ isomer shown in Fig. 2(c) is stabilized by the
attractive hydrogen-chloride interactions.
The geometries of the lowest energy isomers for (NaCl)$_n$, n = 4,7,8,10
are similar to
the geometries of the lowest energy isomers that we find for the respective
(NH$_4$Cl)$_n$ clusters.
For the remaining isomers, n = 5,6, and 9, the geometries of the the
second lowest isomers of (NaCl)$_n$ are similar to those of the lowest
energy isomers of (NH$_4$Cl)$_n$.
In the work of Diefenbach and Martin \cite{dm}, lowest energy isomers
are identified for for several alkali halide clusters.  They find that
the lowest energy isomer geometries are dependent on both the
constituent ions and the potential model used.
It is noteworthy that unlike bulk ammonium chloride crystal,
for which the cesium chloride structure is thermodynamically stable, we
have found no cesium chloride type structures in the clusters.
The cluster size at which the cesium chloride structure begins to appear
represents an interesting problem.

A major thrust of this paper is the development of the parallel
J-walking algorithm with the use of Parallel Virtual Machine (PVM) package.
The ability to maintain the J-walking distributions
dynamically at each required temperature makes it possible to perform
an ergodic Monte Carlo simulation spanning the entire temperature range
of interest.
The PVM J-walking algorithm is designed to take full advantage of a
multi-processor computing environment.
Ergodic simulations spanning the entire temperature
range of interest require the same amount of time as a single Metropolis Monte
Carlo simulation, provided that a separate processor can be assigned to
each process in a PVM algorithm.
PVM is designed to be most efficient when the passing of information
among the different processes is infrequent.
In that respect, PVM is ideally suited for J-walking simulations because
the most frequent jumps are made only once every fifteen passes in the
computing processes.
The PVM J-walking algorithm does not require the
storage of large external configuration files needed in the serial
algorithm, and
ergodic Monte Carlo simulations can be performed on significantly
larger systems than possible previously.

The constant-volume heat capacities for each cluster size have been
determined, revealing two important temperature regions.
The first region, seen most prominently in the (NH$_4$Cl)$_n$, n=4, 6, 8,
and 9 clusters, is the flat region in the 500 - 800 K
vicinity.
This flat region in the heat capacity is apparently a consequence of
competing contributions from rotational motion of the ammonium ions and the
anharmonic vibrational motion in the clusters.
The other feature of the heat capacity curves for the majority
of the clusters is the prominent maximum in the 1000-1200 K
range that we have called a melting peak.
This peak coincides with isomerization to open-type configurations.
Cluster geometries at temperatures below 1000
K are normally dominated by compact structures belonging to potential
energy wells of low lying isomers.

Future work on ammonium chloride clusters will be directed toward a
quantum mechanical treatment of this system.
J-walking has already been extended to quantum systems by incorporating it
into the Fourier path integral Monte Carlo method. \cite{ffd2}
Realistic inclusion of all degrees of freedom in this system cannot be
accomplished using classical mechanics because of the intrinsic
quantum-mechanical nature of the internal vibrations of the ammonium ions.
As an example of the potential importance of quantum contributions,
it can be anticipated that the quantum-mechanical treatment
will have a significant effect on the heat capacity curve
of the (NH$_4$Cl)$_3$ cluster because the differences in zero-point
energies of the isomers can be expected to
rearrange the ordering and change the energy
spacings between the isomers.
Furthermore, significant isotope effects are known in the bulk system,
implying the importance of quantum effects in the
clusters.~\cite{disorder}

\section*{Acknowledgments}
This work was
supported in part by the National Science Foundation under grant numbers
CHE-9411000 and CHE-9203498.
Acknowledgement is made to the Donors of the Petroleum Research Fund of
the American Chemical Society for partial support of this work.
This research was sponsored in part by the Phillips Laboratory, Air Force
Material
Command, USAF, through the use of the MHPCC under cooperative agreement number
F29601-93-2-0001. The views and conclusions contained in this document are
those of the authors and should not be interpreted as necessarily
representing the official policies or endorsements, either expressed or
implied, of Phillips Laboratory or the U.S. Government.
We thank Jim Doll for helpful discussions
and suggestions.

\pagebreak
\begin{table} \centering
\caption{The parameters used in the model potential$^e$}

\vspace{.15in}
\renewcommand{\arraystretch}{1.5}
\begin{tabular}{*{5}{c}}
Pair & $A_{ij}$ & $\alpha _{ij}$ &  $C_6$ & $D_{12}$  \\
\hline
H-H$^{c,d}$ & 1.0162 &  1.9950 & 2.9973& 2021.01 \\
N-N$^b$ & 104.74 & 1.5611 & 25.393 & 0 \\
Cl-Cl$^b$ & 125.55 & 1.7489 & 113.68 & 0 \\
H-N$^a$ & 10.318 & 1.7780 & 8.7229 & 0 \\
H-Cl$^c$ & 0 & 0 & 10.033 & 43884.0 \\
N-Cl$^a$ & 114.22 & 1.6550 & 53.736 & 0 \\
\end{tabular}
\end{table}
\parbox[b]{10in}{$^a$ Combining rules \\
$^b$ Reference \onlinecite{klein} \\ $^c$ Reference \onlinecite{pr1} \\
$^d$ Reference \onlinecite{pr2}
\\$^e$ Units of energy in Hartree and units of length in Bohr}
\pagebreak
\section*{Figure Captions}
\noindent
Figure 1.  Diagram of a sample PVM J-walking process. Boxes on the left-hand
side represent the generating processes, boxes on the right-hand side
represent the computing processes, and arrows indicate the direction in which
configurations are passed from the generating processes.
\\
\\
Figure 2. (NH$_4$Cl)$_3$ (a) is the lowest energy isomer,V = -0.69206 eV,
(c) is the isomer 235
K above (a), (e) is the isomer 635 K above (a), (b) is the transition state
connecting (a) with (c), 2118 K above (a), (d) is the transition connecting
(a) with (e), 2021 K above (a). The energies of the minima are determined
using Eq. (1).  The transition state barriers include intramolecular
contributions from Eq. (2).
\\
\\
Figure 3. All discovered isomers found for (NH$_4$Cl)$_4$,
(a) is the lowest energy
isomer at V = -0.98950 eV, the energies for the remaining isomers are given
relative to the
lowest energy isomer: (b)13485 K, (c) 13844 K,(d) 16590 K, (e) 18535 K,
(f) 18900 K.
\\
\\
Figure 4. The lowest energy isomers for (NH$_4$Cl)$_n$,
n = 5 - 10; (a) n=5, V = ~-1.22880 eV, (b) n=6, V = -1.519387 eV,
(c) n = 7, V = -1.76779 eV, (d) n = 8, V = -2.05161 eV,
(e) n = 9, V = -2.31868 eV, (f) n = 10, V = -2.58333 eV.
\\
\\
Figure 5. $\Delta$V = V$_{n-1}-$V$_{n}$ as a function of $n$, for the
lowest energy isomers.  The gain in binding energy for the even clusters
is generally greater than the gain for the odd clusters.
\\
\\
Figure 6. The potential energy contribution to the
constant-volume heat capacity for (NH$_4$Cl)$_3$ as a function of T.
\\
\\
Figure 7. The potential energy contribution to the
constant-volume heat capacity for (NH$_4$Cl)$_4$ as a function of T.
The maximum at 1100 K is defined to be a melting peak.
\\
\\
Figure 8. The potential energy contribution to the
constant-volume heat capacities for (NH$_4$Cl)$_n$, n = 5 - 10,
where (a) n = 5, (b) n = 6, (c) n = 7, (d) n = 8, (e) n = 9, and (f) n = 10.
\\
\\
Figure 9. (a) and (b) are examples of compact structures of
(NH$_4$Cl)$_5$, with energies of 160 K and 773 K above the lowest-energy
isomer, and (c) and (d) are examples of open-type structures, with
energies 12850 K  and 15079 K above the lowest energy isomer, respectively.
\\
\\
Figure 10. Histograms of isomer distribution for (NH$_4$Cl)$_7$ at
(a) 425 K, (b) 600 K, and (c)  800 K.  The units of potential energy $V$
in these
graphs are degrees K above the potential energy of the lowest energy
isomer.  The variable $N$ is the number configurations in a Monte Carlo
walk whose nearest local minimum on the potential energy surface has
potential
energy $V$.
\\
\\
Figure 11. Upper curve shows the constant-volume heat capacity for
(NH$_4$Cl)$_9$ in the absence of rotational moves, and the lower curve is
the heat capacity with the translational motion of ammonium and chloride
ions excluded. The sum of the two curves changes little explaining
the plateau features in Fig. 8.

\end{document}